\documentclass[runningheads,a4paper]{llncs}

\usepackage{amssymb}
\usepackage{graphicx}
\usepackage{subfigure}
\usepackage{url}
\usepackage{amssymb}
\urldef{\mailsa}\path|{jlrivero,bribeiro}@dei.uc.pt|    
\newcommand{\keywords}[1]{\par\addvspace\baselineskip
\noindent\keywordname\enspace\ignorespaces#1}
\usepackage{fancyvrb}

\begin{document}
	\title{\vspace*{-0.75cm}{\Large Accepted as a Conference Paper for INNS on Big Data 2016}\\
		Attribute Learning for Network Intrusion Detection}

\mainmatter  


\titlerunning{ALNID: Attribute Learning for Network Intrusion Detection. }

%
%
\author{Jorge Luis Rivero P{\'e}rez%
\and Bernardete Ribeiro}
\authorrunning{Rivero P{\'e}rez, JL. and Ribeiro, B.}

\institute{CISUC - Department of Informatics Engineering, University of Coimbra, Portugal \\
\mailsa
}

%
%

\toctitle{Lecture Notes in Computer Science}
\tocauthor{Authors' Instructions}
\maketitle
\begin{abstract}
Network intrusion detection is one of the most visible uses for Big Data analytics. One of the main problems in this application is the constant rise of new attacks. This scenario, characterized by the fact that not enough labeled examples are available for the new classes of attacks is hardly addressed by traditional machine learning approaches. New findings on the capabilities of Zero-Shot learning (ZSL) approach makes it an interesting solution for this problem because it has the ability to classify instances of unseen classes. 
ZSL has inherently two stages: the attribute learning and the inference stage. In this paper we propose a new algorithm for the attribute learning stage of ZSL. The idea is to learn new values for the attributes based on decision trees (DT). Our results show that based on the rules extracted from the DT a better distribution for the attribute values can be found. We also propose an experimental setup for the evaluation of ZSL on network intrusion detection (NID). 
\end{abstract}
\section{Introduction}
Nowadays society is becoming increasingly dependent on the use of computer systems in various fields such as finance, 
security and many aspects of everyday life. On the other hand threats and attacks are increasingly growing. Cyber-Security research area looks at the ability to act pro-actively in order to mitigate or prevent attacks. In that sense Network Intrusion Detection (NID) is one of the (possible) solutions. This task is basically carried out under two approaches: (i) Missuse Detection 
 and (ii) Anomaly Detection. 
These approaches have advantages and disadvantages associated with their suitability to various scenarios \cite{rivero2014}. There are machine learning based solutions to these approaches \cite{sangkatsanee,tsai}. Despite the extensive academic research, their deployment in operational NID environments has been very limited \cite{sommer}. 
On the other hand new attacks are constantly occurring, often as variants of known attacks. 

Traditional machine learning approaches are unable to tackle challenging scenarios in which new classes may appear after the learning stage. This scenario is present many real world situations \cite{eszsl}. Specifically in NID related tasks where one of the main problems is the emergence of new attacks which corresponds with new classes. In that case the detection algorithm should identify the new classes which is important in real environments. 
Recently, there has been an increasing interest in the study of ZSL approaches, which might be a possible solution to this problem 
\cite{akata,dap,eszsl}. 
In this paper we propose an Attribute Learning for Network Intrusion Detection (ALNID) algorithm. We present the preliminary results as an initial step prior to the detection of new attacks on networks. A proposal of the experimental data setup for the application of ZSL in NID is also given.
The proposed attribute learning algorithm can be generalized to many problems. It is simple and based on criteria such as attributes frequency and entropy, achieving to learn new values for the original attributes. The results show a significant improvement in the representation of the attributes for classes. This is encouraging since we expect to achieve higher accuracy in the inference stage of ZSL.
The rest of the paper is organized as follows. In Section II we briefly review the background,  related work on ZSL as well as the attribute learning prior stage. In Section III we describe the ALNID algorithm. In Section IV we present the data preprocessing and propose an experimental data setup for the application of ZSL in NID. In Section V we discuss the results on KDD Cup 99 dataset. Finally, we conclude and address some lines of future work.
\section{Zero-Shot Learning}
ZSL is inspired by the way human beings are capable of identifying new classes when a high-level description is provided. It consists of identifying new classes without training examples, from a high-level description of the new classes (unseen classes) that relate them to classes previously learned during the training (seen classes). This is done by learning the attributes as an intermediate layer that provides semantic information about the classes to classify. ZSL has two stages: the training or attribute learning stage in which knowledge about the attributes is captured, and then the inference stage where this knowledge is used to classify instances among a new set of classes. This approach has been widely applied to images classification tasks 
\cite{akata,dap,eszsl,sun}. 
There are different solutions for both attribute learning and for the inference stage, which models the relationships among features, attributes, and classes in images. Such strategy makes it difficult to apply ZSL to other problems. This, coupled with the need of identify new attacks on computer networks motivated us to development this research.
\subsection{Attribute Learning Stage}
The attribute learning focuses on learning to recognize several properties from objects, which allow to learn new classes based only on their description \cite{eszsl}. Recently there have been applications of automatic recognition of attributes on images \cite{ferrari,lampert,sun}. In \cite{sun} the authors propose the extensive Scene UNderstanding (SUN) database that contains $899$ categories and $130,519$ images. The database is built by human scene classification using Amazon Mechanical Turk (AMT). The authors firstly defined the attributes and then collected votes from AMT evaluating the presence or not of the previously defined attributes on each image. The procedure then  selected or designed several state-of-art features that are potentially useful for scene classification. 
To the best of our knowledge this approach has never been applied on network intrusion detection.
\subsection{Inference Stage}
In the inference stage the predicted attributes are combined to infer the classes. There are basically three approaches \cite{eszsl} for this second stage: k--nearest neighbour (k--NN), probabilistic frameworks \cite{dap} and energy function \cite{akata,eszsl}. 
One interesting technique is the cascaded probabilistic framework proposed in \cite{dap,lampert} where the predicted attributes on the first stage are combined to determine the most likely target class. It has two variants: the Directed Attribute Prediction (DAP), in which during the training stage, for each attribute, a probabilistic classifier is learned. Then, at inference stage the classifiers are used to infer new classes from their attributes signatures. 
The other variant is the Indirected Attribute Prediction (IAP) which learns a classifier for each training class and then combines predictions of training and test classes. The third ZSL approach uses energy function \cite{akata,eszsl}, which is the same as the penalty function where given $ x $: data, and $v$: category vector, the energy function $ E_w (x,v) = x'Wv $ is trained, with the metric $ W $ being $ E_w (x,v) $ positive if $ x=v $  and negative if $ x!=v $. 
In \cite{crossmodal} a framework is proposed to predict both seen and unseen classes. Unlike other approaches this proposal not only classifies unknown classes but also classifies known classes. In the attribute learning stage they consider the set of all classes $Y$ during training and testing. Some classes $y$ are available as seen classes in training data ($Y_s$) and the others are the Zero-Shot classes, without any training data, as unseen classes ($Y_u$). Then they define $W = W_s  \cup W_u$ as the word vectors for both, seen and unseen classes. All training images $x^{(i)}  \in X_y $ of a seen class $y \in Y_s$ are mapped to the word vector $w_y$. Then, to train this mapping a two-layer neural network to minimize the following objective function (Equation \ref{eq:neuralnetwork}) is trained \cite{crossmodal}:
\begin{equation}
J(\Theta ) = \sum\limits_{y \in Y_s }^{} {\sum\limits_{x^{(i)}  \in X_y }^{} {\left\| {w_y  - \theta ^{(2)} f(\theta ^{(1)} x^{(i)} )} \right\|^2 } }
\label{eq:neuralnetwork} 
\end{equation}
Other recent and simple energy function based approach is the ESZSL proposal \cite{eszsl}. It is based on \cite{akata} which models the relationships among features, attributes, and classes. 
The authors assume that at training stage there are $z$ classes, which have a signature composed of $a$ attributes. That signatures are represented in a matrix $ S \in [0,1]^{axz} $. This matrix contains for each attribute any value in $[0, 1]$ representing the relationship between each attribute and the classes. However, how the matrix $S$ is computed is not addressed. The instances available at training stage are denoted by
$X \in \mathbb{R}^{dxm} $, 
where $d$ is the dimensionality of the data, and $m$ is the number of instances. The authors also compute the matrix $ Y \in \{  - 1,1\} ^{mxz} $ to denote to which class belongs each instance. During the training they compute the matrix 
$V \in \mathbb{R}^{dxa} $ as in Equation \ref{eq:eszsl}:
\begin{equation}
V = (XX^T  + \gamma I)^{ - 1} XYS(SS^T  + \lambda I)^{ - 1} 
\label{eq:eszsl} 
\end{equation}
In inference stage they distinguish between a new set of $z'$ classes by their attributes signatures, $ S' \in [0,1]^{axz'} $. Then, given a new instance $x$ the prediction is given by Equation \ref{eq:3}: 
\begin{equation}
\mathop {\arg \max }\limits_i (x^T VS_i ')
\label{eq:3} 
\end{equation}
\begin{figure}[t!]
	\centering
	\includegraphics[keepaspectratio, width= 8.0cm]{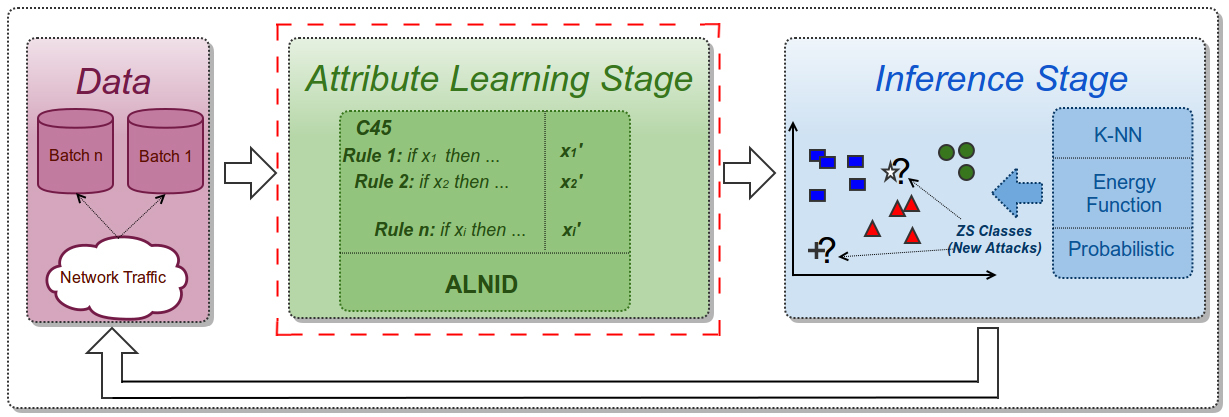}
	\caption{Zero-Shot Learning for Network Intrusion Detection}
	\label{fig:zero-shot}
\end{figure}
\section{Proposed Attribute Learning Algorithm}
Most of the attribute learning algorithms have been implemented for automatic recognition of attributes on images based on the features extraction \cite{ferrari,lampert,sun}. It is hardly applicable to other kind of problems like NID. Firstly we evaluate different machine learning algorithms on the preprocessed dataset (which is described in next section) and the C45 decision tree algorithm  shows the higher classification accuracy of 99.54\%. This result together the lack of attribute learning algorithms for non computer vision related tasks leads us to think in the extracted rules as a solution to build or relearn the attributes. This method could be extended not only to cover a wider area of Information Technology security but also to different kind of problems with labeled and structured data in which the C45 algorithm could be applied with good results.

Our proposal is for the Attribute Learning Stage of the ZSL approach depicted in Figure \ref{fig:zero-shot}. ALNID - Attribute Learning for Network Intrusion Detection Algorithm - is a rule-based algorithm which weights the attributes according its entropy and frequency in the rules. We begin with a set of $X$ instances composed of $ A = \{ a_1 ,\cdots,a_n \} $ attributes. For each $a_n  \in A$ we compute the quantity of information ($I$), the entropy ($E$) and the information gain ($G$). With these values we compute the C45 algorithm for decision trees. 
Furthermore, during each iteration we record how many times each $a_i  \in A$ attribute is evaluated by each rule of the set $ R = \{ r_1 ,\cdots,r_m \} $. This is what we call frequency. Then, a new set of attributes $A' = \{ a_1 ',\cdots,a_n '\} $ is created. The values of $A'$ are the frequency count, increasing by each time that an attribute is evaluated by the rule $r_m$. As output the algorithm returns the set of new valued instances $X' = \{ A_i '\} _{i = 1}^N$ composed of the learned attributes, where $N$ is the number of instances. The pseudo-code of the proposed method is listed below:
\begin{verbatim}
{attributes weighted w.r.t. their frequency in each rule match};
X: Examples, A: Attributes (Input)
X': Examples with learned attributes (Output) 
var
    I: Quantity of information
    E: Entropy
    G: Information Gain
begin
    if all examples are from the same class then
        repeat
            Weight attribute;
            Add examples to X' 
        until attribute in A    
        return X';
     else
         I = Compute quantity of information of the examples;
         repeat
             E = Compute the entropy of attribute;
             G = Compute the information gain of attribute;
         until attribute in A
         a = attribute that maximizes G;
         v = value of a; 
         Delete a from A;
         Generate a root node for the attribute a;
         Weight attribute;
         Add examples to X'
         repeat
             ALNID (examples from X with a == v, A);
             generate a new branch with a==v;
         until partition in Partitions generated by values of attribute a      
         return X';
end.
\end{verbatim}
\begin{table}[htb]
	\centering \small
	\caption{Data Setup for Zero-Shot Learning on Network Intrusion Detection}
	\begin{tabular}{lccr}
		\hline\noalign{\smallskip}
		\textbf{Class} & \textbf{Nr. of Examples} & \textbf{Zero-Shot Class} & \textbf{Category}\\
		\noalign{\smallskip}
		\hline
		\noalign{\smallskip}
		smurf & 280,790 & no & DOS (D) \\
		neptune & 107,201 & no & DOS (D) \\
		back & 2203 & no & DOS (D)\\
		teardrop & 979 & yes & DOS (D)\\
		pod & 264 & no & DOS (D)\\
		land & 21 & yes & DOS (D)\\
		normal & 97,277 & no & NORMAL (N) \\
		satan & 1589 & no & PROBE (P) \\
		ipsweep & 1247 & yes & PROBE (P) \\
		portsweep & 1040 & no & PROBE (P) \\
		nmap & 231 & yes & PROBE (P) \\
		warezclient & 1020 & no & R2L (R) \\
		guess\_passwd & 53 & yes & R2L (R) \\
		warezmaster & 20 & no & R2L (R) \\
		imap & 12 & yes & R2L (R) \\
		ftp\_write & 8 & no & R2L (R) \\
		multihop & 7 & no & R2L (R) \\
		phf & 4 & no & R2L (R) \\
		spy & 2 & no & R2L (R) \\
		buffer\_overflow & 30 & no & U2R (U) \\
		rootkit & 10 & yes & U2R (U) \\
		loadmodule & 9 & no & U2R (U) \\
		perl & 3 & yes & U2R (U) \\	    		
		\hline
	\end{tabular}
	\label{table:Table2}
\end{table}
\section{Data and Experimental Setup}
KDD Cup 99\footnote{KDD Cup is the annual Data Mining and Knowledge Discovery competition organized by ACM Special Interest Group on Knowledge Discovery and Data Mining.}
is the most used in research as data to test and compare intrusion detection algorithms.
It contains about 5 millions of instances representing 39 types of attacks which are grouped into 4 categories and also one category for normal traffic. Some preprocessing variants applied on the dataset are based on the reduction of the classes replacing the attack labels by their corresponding categories, thus reducing the number of classes to 5 \cite{rivero2014}. Each instance represents a TCP/IP connection composed of 41 attributes both quantitative and qualitative 
\footnote{Current research uses a small portion that represents the $10\%$ of the original dataset containing $494,021$ instances.}. 
Despite our proposal holds for the attribute learning stage, we propose herein a scheme to apply ZSL approach in NID tasks (see Figure \ref{fig:zero-shot}). 
Table \ref{table:Table2} shows that classes such as \emph{spy} and \emph{perl} have 2 and 3 instances respectively while classes as \emph{smurf} and \emph{normal} have 280,790 and 97,277 respectively. Then, considering the study in \cite{rivero2014} we selected the 12 attributes from the original 41 ones. Its statistical description are shown in Table \ref{table:Table3}. We modified the dataset using the five categories as classes. Later, for each category we selected two attacks as Zero-Shot classes.  
Table \ref{table:Table2} illustrates these values and the Zero-Shot classes. This setup is very practical for the application at hand because we can classify the Zero-Shot classes. In this case, new attacks can be classified in the categories to which they belong to. 
\section{Results and Discussion}
\begin{figure}
	\centering
	\subfigure[srv\_count]{\includegraphics[width=0.35\textwidth]{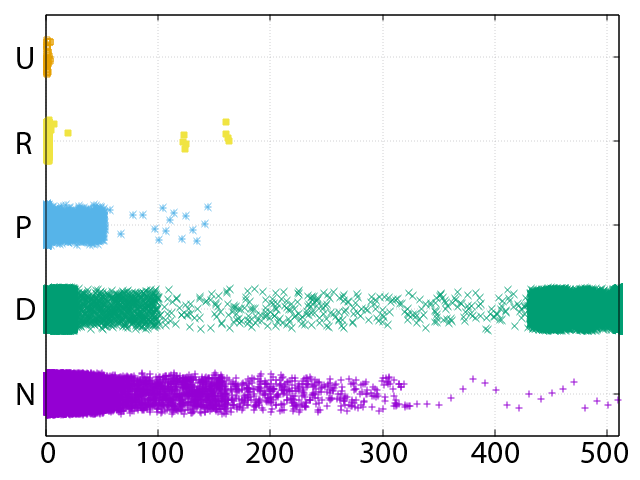} }	
	\subfigure[\emph{srv\_count'}]{\includegraphics[width=0.35\textwidth]{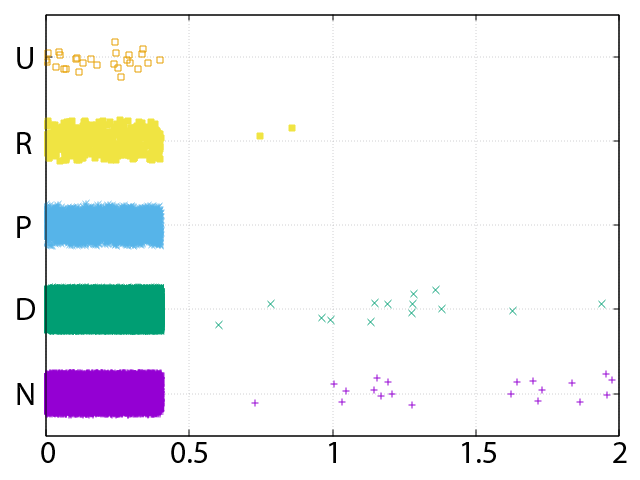} }
	\subfigure[sam\_srv\_rate]{\includegraphics[width=0.35\textwidth]{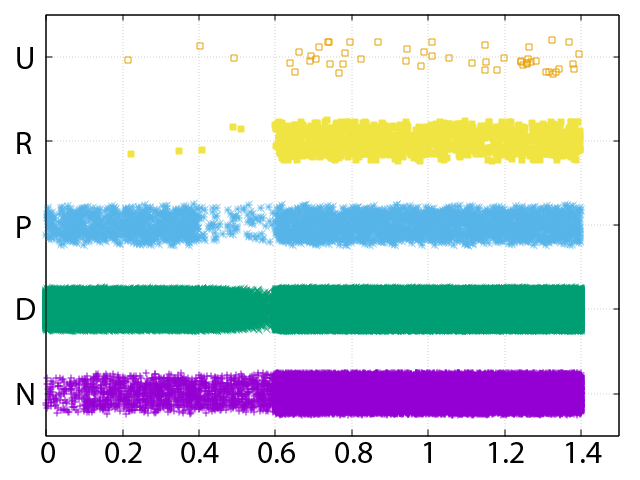} }	
	\subfigure[\emph{sam\_srv\_rate'}]{\includegraphics[width=0.35\textwidth]{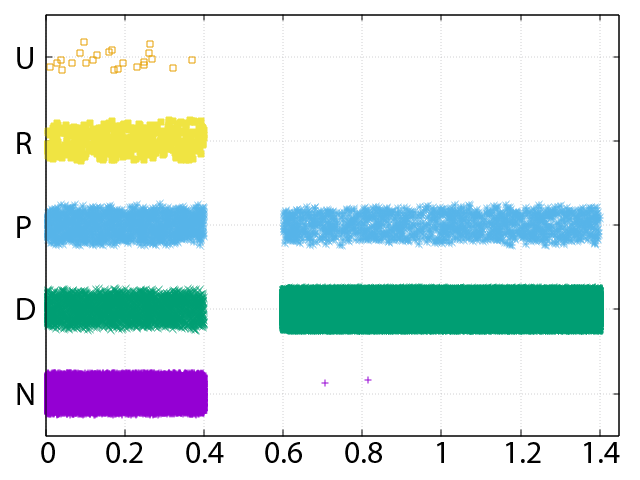} }
	\subfigure[dst\_host\_count]{\includegraphics[width=0.35\textwidth]{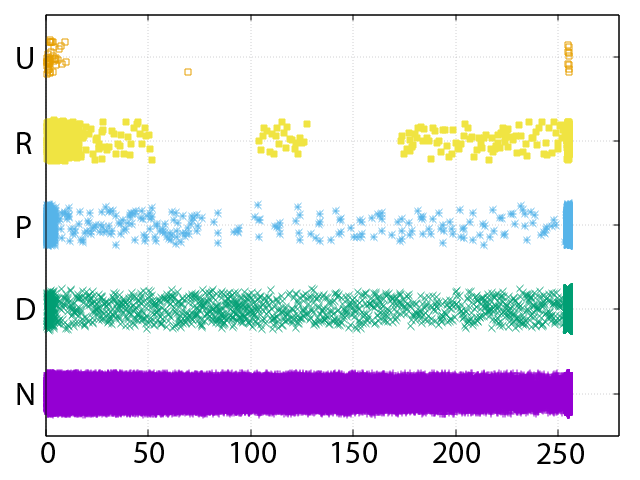} }	
	\subfigure[\emph{dst\_host\_count'}]{\includegraphics[width=0.35\textwidth]{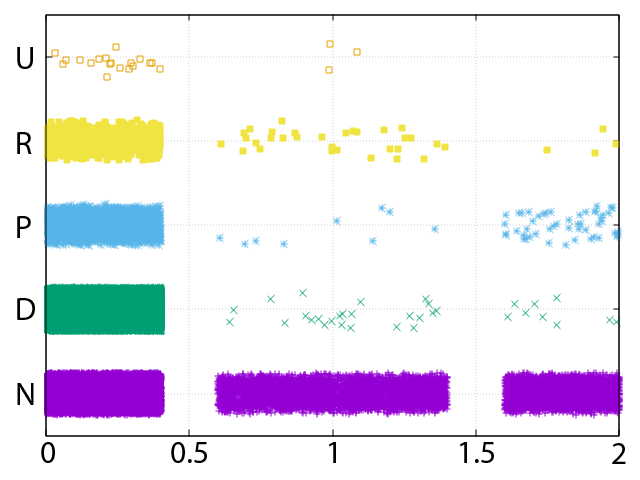} }
	\subfigure[dst\_host\_srv\_count]{\includegraphics[width=0.35\textwidth]{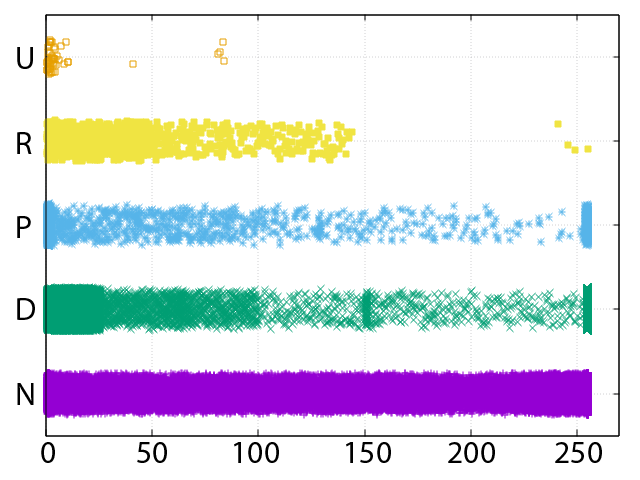} }	
	\subfigure[\emph{dst\_host\_srv\_count'}]{\includegraphics[width=0.35\textwidth]{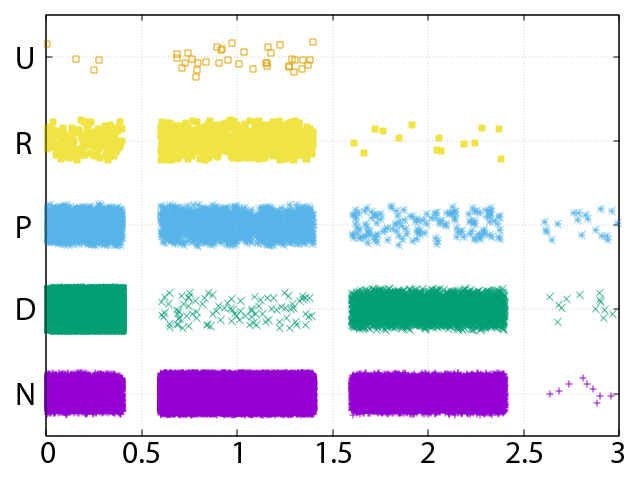} }
	\subfigure[dst\_host\_same\_srv\_rate]{\includegraphics[width=0.35\textwidth]{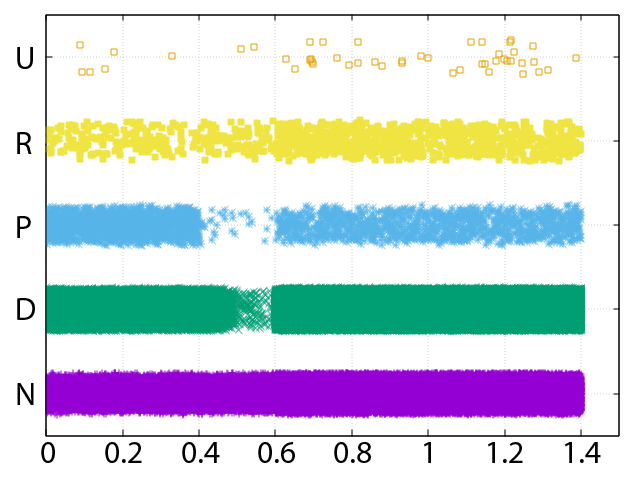} }	
	\subfigure[\emph{dst\_host\_same\_srv\_rate'}]{\includegraphics[width=0.35\textwidth]{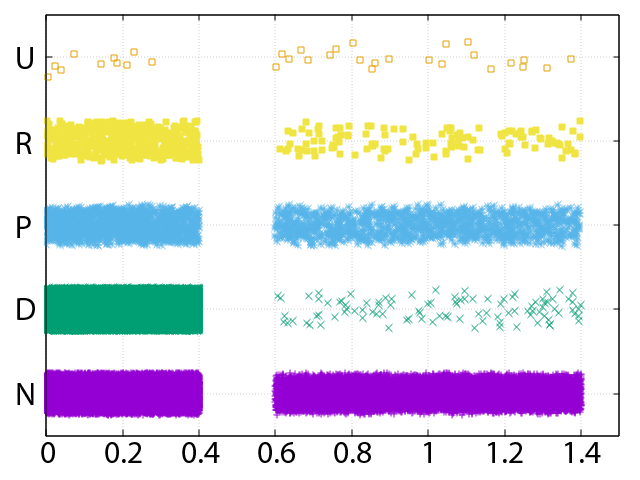} }
	\subfigure[dst\_host\_same\_port\_rate]{\includegraphics[width=0.35\textwidth]{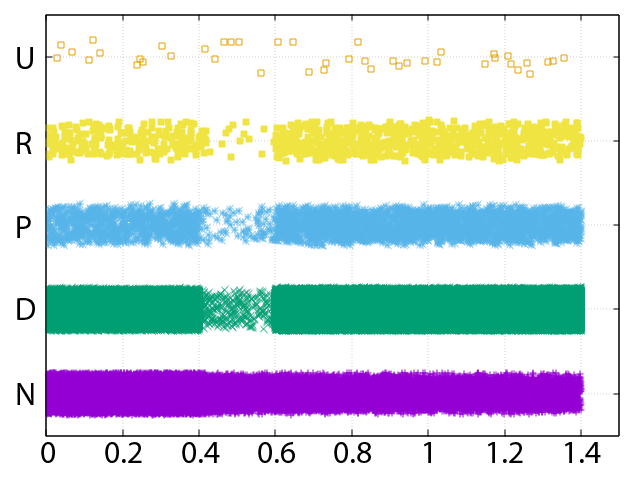} }	
	\subfigure[\emph{dst\_host\_same\_port\_rate'}]{\includegraphics[width=0.35\textwidth]{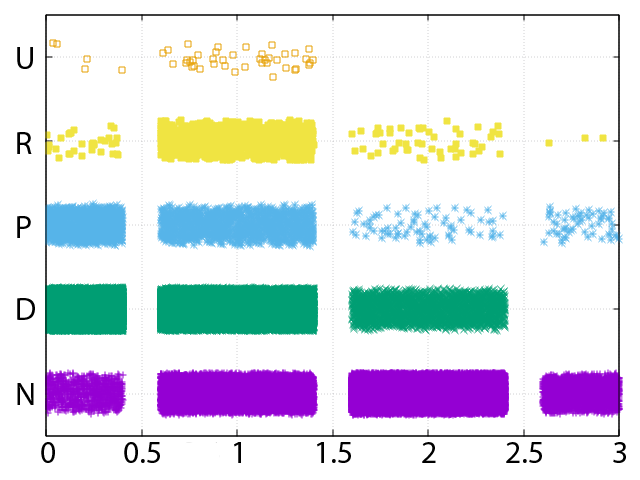} }
	\caption{Original (left) and learned (right) attributes. }	
	\label{fig:two}	
\end{figure}

The ALNID algorithm was evaluated on the preprocessed dataset computing a DT of 132 leaves and 263 rules ($ R = \{ r_1 ,...,r_{263} \} $). The classification accuracy for the seen classes was 99.94\%. 
\begin{table}[htb]
	\centering \small
	\caption{Statistical description of the attributes}
	\begin{tabular}{lcccr}
		\hline\noalign{\smallskip}
		\textbf{\small{Attributes}} & \textbf{\small{Minimum}} & \textbf{\small{Maximum}} & \textbf{\small{Mean}} &  \textbf{\small{StdDev}} \\
		\hline
		\noalign{\smallskip}
		duration & 0 & 58,329 & 47.979 & 707.747 \\
		\emph{duration'} & 0 & 3 & 0.013 & 0.117 \\
		\hline		
		protocol\_type & 1 & 3 & 2.189 & 0.961 \\
		\emph{protocol\_type'} & 0 & 1 & 0.845 & 0.362 \\
		\hline		
		src\_bytes & 0 & 693,375,640 & 3025.616 & 988,219.101 \\
		\emph{src\_bytes'} & 0 & 7 & 0.762 & 1.526 \\
		\hline		
		dst\_bytes & 0 & 5155,468 & 868.531 & 33,040.035 \\
		\emph{dst\_bytes'} & 0 & 3 & 0.057 & 0.297 \\
		\hline		
		urgent & 0 & 3 & 0 & 0.006 \\
		\emph{urgent'} & 0 & 1 & 0 & 0.016 \\
		\hline		
		count & 0 & 511 & 332.286 & 213.147 \\
		\emph{count'} & 1 & 4 & 1.821 & 0.425 \\
		\hline		
		srv\_count & 0 & 511 & 292.907 & 246.323 \\
		\emph{srv\_count'} & 0 & 2 & 0 & 0.017 \\
		\hline		
		same\_srv\_rate & 0 & 1 & 0.792 & 0.388 \\
		\emph{same\_srv\_rate'} & 0 & 1 & 0.784 & 0.411 \\
		\hline		
		dst\_host\_count & 0 & 255 & 232.471 & 64.745 \\
		\emph{dst\_host\_count'} & 0 & 2 & 0.033 & 0.238\\
		\hline		
		dst\_host\_srv\_count & 0 & 255 & 188.666 & 106.04 \\
		\emph{dst\_host\_srv\_count'} & 0 & 3 & 0.21 & 0.469 \\
		\hline		
		dst\_host\_same\_srv\_rate & 0 & 1 & 0.754 & 0.411\\
		\emph{dst\_host\_same\_srv\_rate'} & 0 & 1 & 0.024 & 0.154 \\
		\hline		
		dst\_host\_same\_src\_port\_rate & 0 & 1 & 0.602 & 0.481 \\	    
		\emph{dst\_host\_same\_src\_port\_rate'} & 0 & 3 & 0.585 & 0.779 \\
		\hline
	\end{tabular}
	\label{table:Table3}
\end{table}
Table \ref{table:Table3} summarizes the minimum, maximum, mean and standard deviation values for each of the original attributes and the learned ones by our proposal.  
The Figure~\ref{fig:two} depicts the distribution per class of the original attribute values ($A = \{ a_1,...,a_{12}\} $) and the learned ones ($A' = \{ a_1 ',...,a_{12} '\} $). These were listed by decreasing order of the entropy ($E$) of the learned attributes. In general the graphical plots show better separated distributions per class for the learned attributes with our proposal than for the originals ones. The learned attribute \emph{duration'} in Figure \ref {fig:two}(\emph{a}) and (\emph{b}) improves distribution at least w.r.t. one class, e.g. the NORMAL (N) one. The rest of the learned attributes by ALNID achieves a higher separability than the original ones in their distribution regarding the classes (see Figure \ref {fig:two}(\emph{c})--(\emph{x})). This new representation of the learned attributes are expected to improve the  k--NN classification during the inference stage.
\begin{figure}
\centering
	\subfigure[duration]{\includegraphics[width=0.35\textwidth]{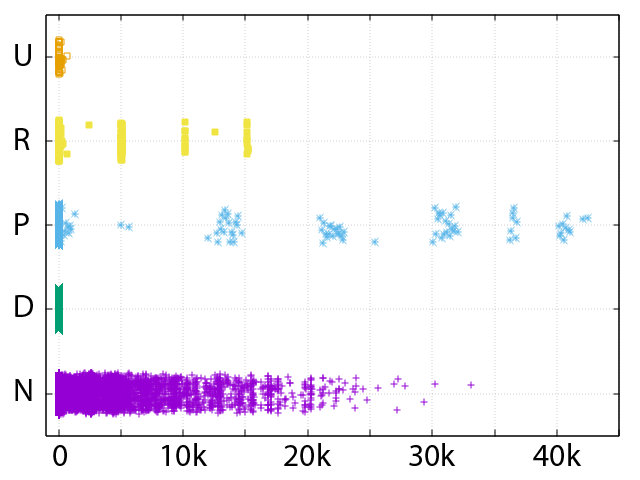} }	
	\subfigure[\emph{duration'}]{\includegraphics[width=0.35\textwidth]{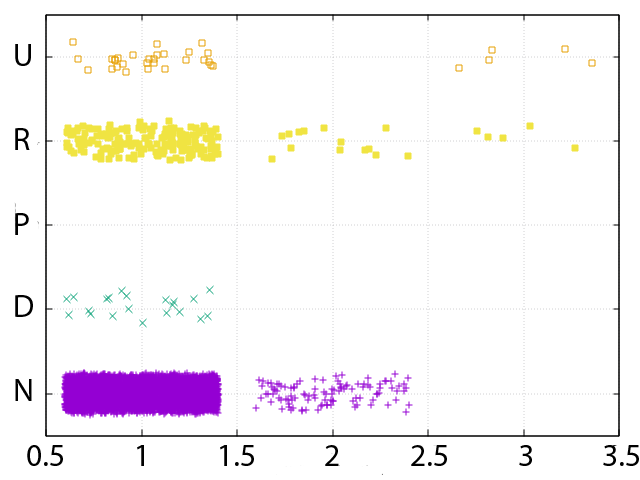} }
	\subfigure[protocol\_type]{\includegraphics[width=0.35\textwidth]{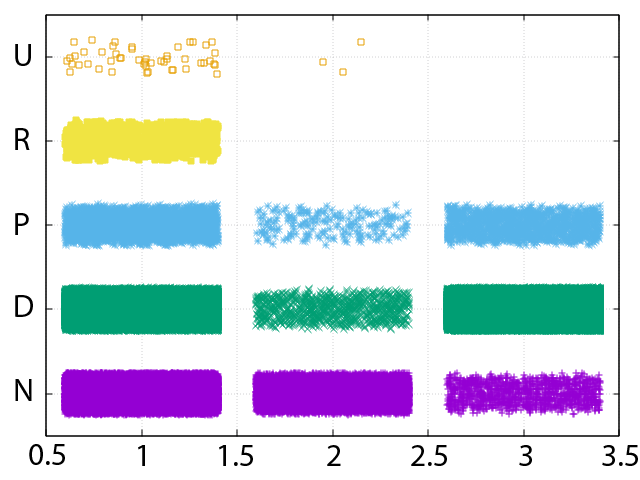} }	
	\subfigure[\emph{protocol\_type'}]{\includegraphics[width=0.35\textwidth]{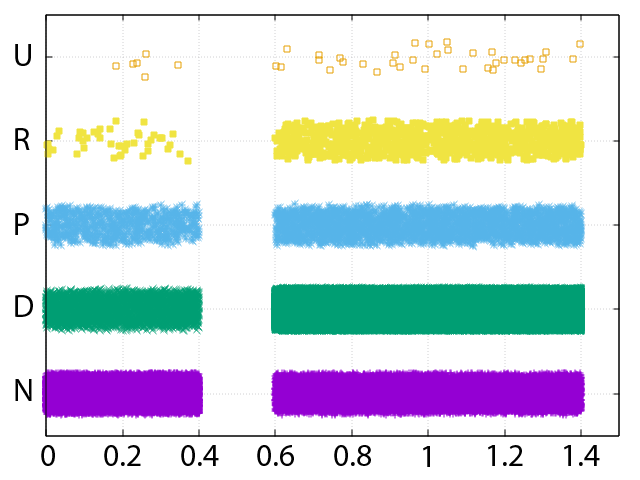} }
	\subfigure[src\_bytes]{\includegraphics[width=0.35\textwidth]{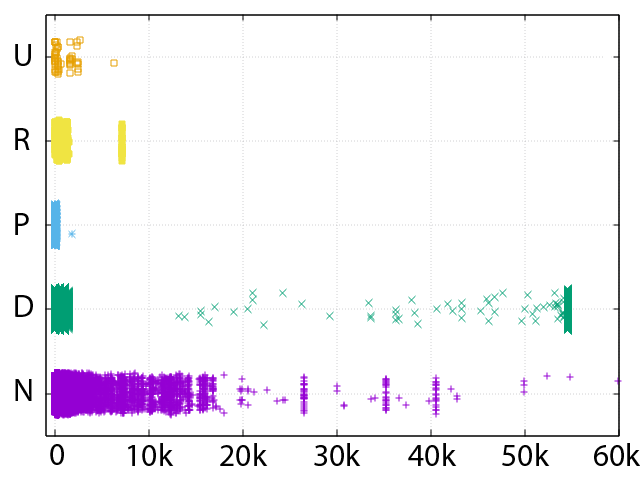} }	
	\subfigure[\emph{src\_bytes'}]{\includegraphics[width=0.35\textwidth]{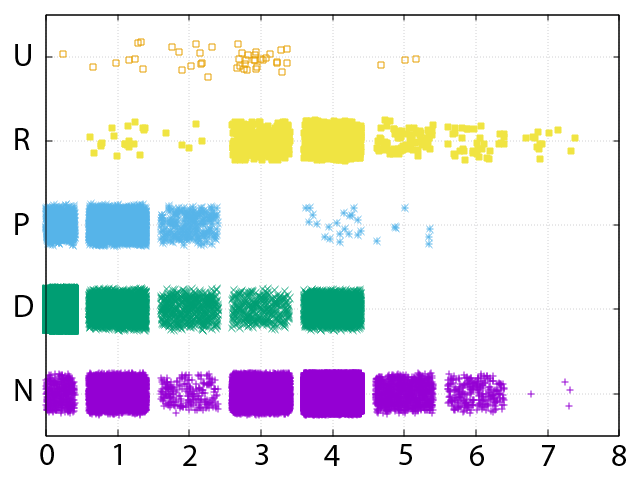} }
	\subfigure[dst\_bytes]{\includegraphics[width=0.35\textwidth]{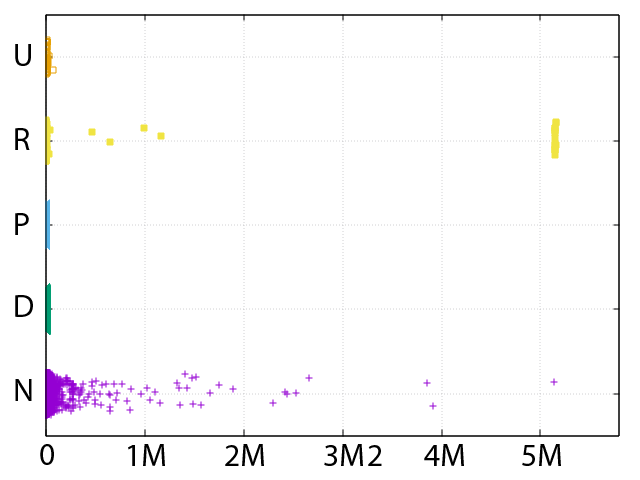} }	
	\subfigure[\emph{dst\_bytes'}]{\includegraphics[width=0.35\textwidth]{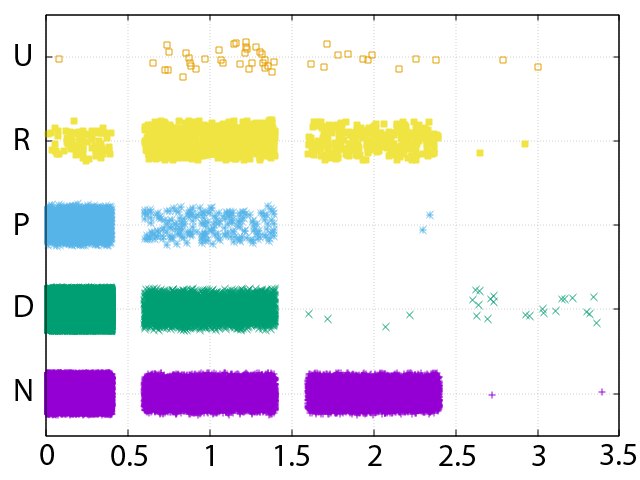} }
	\subfigure[urgent]{\includegraphics[width=0.35\textwidth]{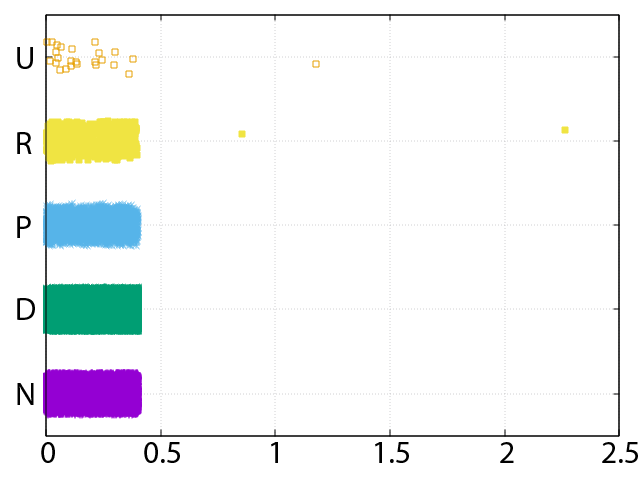} }	
	\subfigure[\emph{urgent'}]{\includegraphics[width=0.35\textwidth]{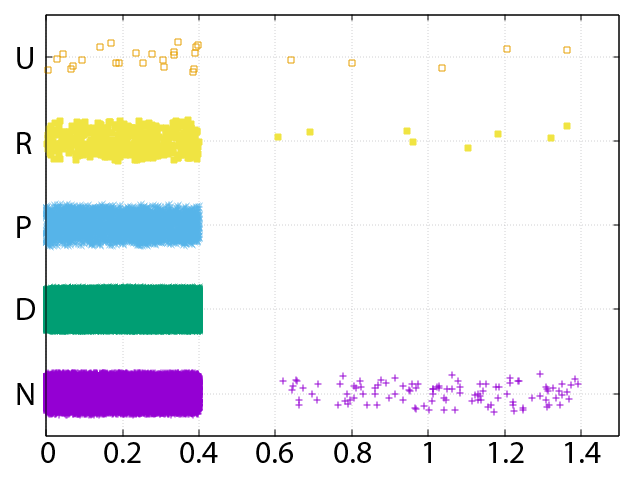} }
	\subfigure[count]{\includegraphics[width=0.35\textwidth]{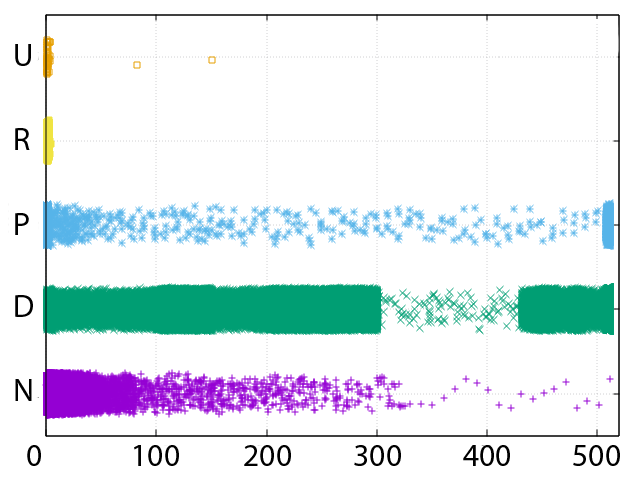} }	
	\subfigure[\emph{count'}]{\includegraphics[width=0.35\textwidth]{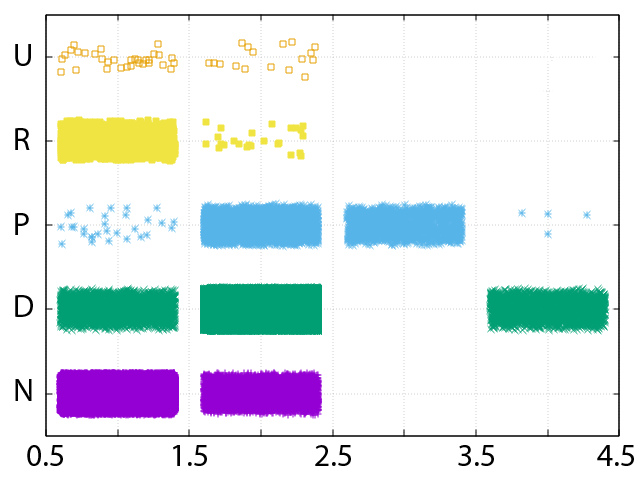} }	
	\label{fig:one}	
\end{figure}
Also we found a way to compute the signature matrix $ S \in [0,1]^{axz} $ required by the energy function used in the inference approach in \cite{eszsl}. The range of values of the learned attributes is discrete which can easily be integrated in the inference stage. 

\section{Conclusions}
ZSL is a two-stage approach that addresses the classification of new classes without any example during the training. 
This, joined with the need to detect new attacks on traffic networks motivated us to research in the application of ZSL to NID. In this paper we proposed ALNID, a new algorithm for the attribute learning stage of ZSL. The algorithm builds a DT and combines entropy and frequency of the original attributes in a weighted function setting. 
Our evaluation proposal on KDD Cup 99 dataset showed better distribution of the attribute values per class. The class separability is convenient for the inference stage based on k--NN. 
Future work will extend the work to the inference stage based on the learned attributes by this proposal and will validate ZSL to other  NID Big Data.

\bibliographystyle{splncs}
\bibliography{zsl}

\end{document}